\documentclass[10pt]{article}
\usepackage[letterpaper]{geometry}
\usepackage{hicss}
\usepackage{times}
\usepackage[none]{hyphenat}
\usepackage{latexsym}
\usepackage{indentfirst}
\usepackage{graphicx}
\graphicspath{{images/}}
\usepackage[hyphens]{url}
\usepackage{booktabs} 
\usepackage{amsmath}
\usepackage{multirow}
\usepackage{amssymb}
\usepackage{float}
\usepackage[normalem]{ulem}
\usepackage[T1]{fontenc}
\usepackage{subcaption}
\usepackage{comment}

\title{``Don’t Downvote A\$\$\$\$\$\$s!!'': An Exploration of Reddit’s Advice Communities}

\author{
    Emily Cannon \\
    School of Information \\
    San Jose State University\\
    {\underline{emily.cannon@sjsu.edu}} \\
    \And
    Bianca Crouse \\
    School of Information \\
    San Jose State University\\\
    {\underline{ bianca.crouse@sjsu.edu}}\\
    \And 
    Souvick Ghosh \\
    School of Information \\
    San Jose State University \\
    {\underline{souvick.ghosh@sjsu.edu}} \\
    \AND
    Nicholas Rihn \\
    School of Information \\
    San Jose State University \\
    {\underline{nicholasrihn@gmail.com}}\\
    \And
    Kristen Chua\\
    School of Information \\
    San Jose State University\\
    {\underline{kristen.chua@sjsu.edu}}\\
    }

\date{}

\begin{document}

\maketitle

\begin{abstract}
Advice forums are a crowdsourced way to reinforce cultural norms and moral behavior. Sites like Reddit contain massive amounts of natural language human interaction, with rules and norms unique to each individual subreddit community. To explore this data, we created a dataset with top 1000 posts from each of two such forums, r/AmItheAsshole and r/relationships, and extracted natural language features including sentiment, similarity, word frequency, and demographics using both algorithmic and manual methods. 
Further, we developed a method to extract demographic information from the subreddits, examined how the post authors’ self-disclosures reflect the unique communities in which their posts are shared, and discussed how the authors’ language use choices might be related to broader social patterns.
We observed some differences between the subreddits in terms of word frequency, demographics disclosure, and gendered language. In general, both subreddits had more female posters than male, and posters tended to use more words about their opposite gender than the same. Gender-diverse posters were uncommon. 
Implications for future research include a more careful, inclusive focus on identity and disclosure and how that interacts with advice-seeking behavior in online communities.
\end{abstract}



\section{Introduction}
Over the last 20 years, social media has become an indispensable part of modern civilization. While sites like Facebook and Twitter emphasize personal connections and networks, others offer a space for content creation, information sharing, and discussion, both private and public, around shared interests. Users of these sites often use a pseudonym and may take care to conceal their real-life identities, but nonetheless participate in communities that both reflect cultural norms and affect society at large. 

Advice forums may fall into a few categories. Some, like Stack Exchange, offer more of a pragmatic how-to or problem-solving service, with an emphasis on topics like programming and IT. The information exchanged rarely requires personal disclosure outside what may be relevant context to address technical issues. Others, like the recently shut Yahoo! Answers, see a broad range of factual and opinion-based inquiries, but offer only superficial and often counterfactual results, without any moderation or filtration. Reddit, a popular social news and discussion site founded in 2005, hosts over 4,000 topical forums, or subreddits, each with its own  rules, conventions, participants, and moderation team. Popular forums range from the practical (r/DIY) to the political (r/worldnews), from ask an expert (r/IamA) to sharing everyday tips and tricks (r/LifeProTips). Each subreddit offers not only a space to share and learn about a particular topic, but a community built around a shared affinity or attitude. Regular participants, through their pseudononymous profiles, often get to know each other and reinforce the culture of that community.

Reddit's advice forums, such as r/relationships
(3 million subscribers) and r/AmItheAsshole
(2.4 million subscribers), offer semi-structured discussions around social interactions that also reflect -- and contribute to -- discourse off the site, as posts are shared and discussed widely on other social platforms (like Twitter). These forums provide posters with an opportunity to provide a uniquely high level of personal disclosure while remaining anonymous. The r/relationships and r/AmItheAsshole subreddits were chosen since they are frequently the source of ``best-of'' listicles~\cite{watson2020} and other media coverage, exploring themes around sexism~\cite{valenti2020} and politics~\cite{martin2017}, as well as scholarly research~\cite{botzer2021}. We are interested in analyzing the users, interaction metrics, and text of some of these forums, as well as drawing distinctions between them. 

Based on our observations of the target subreddits, we 
predicted that most users requesting advice are in their 20s and 30s and are evenly split between men and women. Although in general male users outnumber female users on Reddit by about 2:1~\cite{Pew2019}, we believe the demographics are more evenly split on advice subreddits because they appear to be more popular with women. 
The lower number of posts by males seeking advice can be attributed to the societal definitions of masculinity and advice seeking which is considered a predominantly female genre both historically as well as academically~\cite{neville2012you}. 
Also, posts have more interaction (upvotes, downvotes, comments, rewards) if they are ``entertaining'' in some way. For example, if the post is outlandish or if the advice requester is arguing with the commenters.

Advice forums reinforce cultural norms and moral decision-making in a crowdsourced fashion. 
Forum posters can appeal to the wisdom of the crowd and find themselves redeemed or appropriately chastened, while all participants and observers gain new insight into ``good'' and ``bad'' behavior. What biases might be inherent in this system? Who seeks advice from such a forum, and what do they hope to learn from it? What is it about the advice-seeking itself that garners community input? By answering these questions, we can gain valuable insights into how users interact with social media platforms. 

In this paper, we have explored these questions by applying various natural language processing techniques on Reddit post data. 
We created a dataset containing the natural language of Reddit forums, which contain massive amounts of genuine interactions that are largely uninfluenced by third party stakeholders. 
The main contribution of this paper is an exploration of how natural language processing tools and techniques can be used to examine Reddit, whether those examinations are fruitful, and limitations and opportunities with this type of data. Our research questions could be summarized as follows:

\noindent
\textbf{RQ1:} What natural language features can we extract from unstructured Reddit posts?

\noindent
\textbf{RQ2:} How do online communities with specific and similar interests compare in terms of their natural language features, such as sentiment scores and high frequency words?

\noindent
\textbf{RQ3:} Is there any correlation between a post’s natural language features and its engagement metrics?

%

The Introduction (Section 1) is followed by Section 2 which describes current research using Reddit data and explores related psychological and Human-Computer Interaction (HCI) works. Next, Section 3 includes the description of the data collected. Section 4 explains the feature generation process and the machine learning algorithms used for analysis. Section 5 discusses the results of our analysis while Section 6 provides an overall summary of the work done, points out the limitations and provides direction for future research.


\section{Background and Related Works}

The value of social media sites as an information resource is well established. 
On a surface level, various social media sites including Facebook, Twitter, Reddit, and numerous others appear relatively similar. However, each site contains  information flows that dictate the depth, truthfulness, and originality of interactions. Cinelli et al. (2021)~\cite{cinelli2021echo} explored the frequency of echo chambers across different social media platforms, finding that Facebook and Twitter are more prone to siloed information exchanges. Choi et al. (2016)~\cite{choi2016} compared social media sites to learn their individual qualities as resources for analysis. They measured frequency of posting, presence of external links, and uniqueness of word usage. These types of analysis provide a framework for selecting the appropriate social media site for a given research question.

Among the major social media platforms, Reddit is a particularly rich resource. While Reddit communication remains entirely digital, Nevard (2018)~\cite{nevard2018digital} suggested that Reddit is a public sphere that extends beyond the boundaries of the Reddit environment and influences individuals as they interact with society at large. 
Yadav et al. (2021)~\cite{yadav2021} highlighted the absence of marketing schemes on Reddit, as individual users can report spam. This feature of Reddit ensures the genuineness of the individuals by reducing incentives for profit and discouraging misaligned motivations. Another key feature of Reddit is the presence of metadata provided by the author of the posts. This often appears in a shorthand format that can be extracted using natural language processing tools (Haworth et al., 2021)~\cite{haworth2021classifying}. Collisson et al. (2018)~\cite{collisson2018should} highlighted this feature, and concluded that Reddit ``seems to be an ideal source of real-world, dyadic data from people seeking and offering relationship advice'' (Collission et al., 2018, p. 302).

Reddit is built upon a vast and diverse ecosystem of subreddits. The benefits of Reddit as a data source are common across many of these subreddits, and it is common for researchers to draw on individual subreddits to collect social media data that fits the criteria for their work.  For example, Collisson et al. (2018)~\cite{collisson2018should} specifically examined the subreddit called r/relationships where posters seek individualized relationship advice. The subreddit r/AmItheAsshole is an immensely valuable expression of human morality and ethics that proves useful across multiple domains of study, including gender, psychology, ethics and morality, human computer interaction, and sociology. O'Brien (2020)~\cite{obrien2020dataset} created a public dataset of r/AmItheAsshole posts to support such explorations.

Studies examining gender differences on Reddit endeavor to profile the experiences and behaviors of different gender groups across the site. This type of profiling could lead to improved understanding of gender dynamics transferable to a broad range of social interactions. Thelwall \& Stuart (2019)~\cite{thelwall2018shes} found gender differences in participation rates across different subreddits. Flesch (2019)~\cite{flesch2019mapping} established the practice of collecting gender information from  posters' self-declaration, and confirmed the value of extracting linguistic features from the Reddit corpora to understand gender differences as they relate to different discussion themes. While these works capture gender participation within the Reddit ecosystem, Li et al. (2019)~\cite{liSiyue2019Csso} replicated established gender theories in anonymous digital communities. Their results showed a connection between emotional disclosures and assumptions of female identity.

The anonymity present on Reddit creates a unique space where posters are often willing to disclose personal experiences more candidly than normal. This creates a vast collection of personal accounts that psychologists can use to research the human mind using unobtrusive methods. Curiskis et al. (2007)~\cite{curiskis2020} studied data processing techniques to improve the usability of such data, because a careful analysis of linguistic features can reveal significant insight into the cognition and emotion of both posters and commenters~\cite{collisson2018should, zhou2021assessing, davis2021}. Jaerch et al. (2015)~\cite{jaech2015talking} investigated the relationship between user reputation and engagement on the site. The possibility of transferring results across subreddits is a persistent theme in the literature. Most notably, Botzer et al. (2021)~\cite{botzer2021} classified the moral judgements of r/AmItheAsshole, then applied that knowledge to numerous other subreddits. Their results revealed characteristics of individual posters whose behavior is deemed to have a negative morale valence.
Botzer et al.’s (2021)~\cite{botzer2021} work is relevant from a psychological perspective, but also as an example of the value of Reddit for designing artificial intelligence. The r/AmItheAsshole subreddit asks participants to tag the original posts as a judgement of the original poster's behavior. In addition to Botzer et al. (2021)~\cite{botzer2021}, Zhou et al. (2021)~\cite{zhou2021assessing} and Lourie et al. (2021)~\cite{lourie2020scruples} utilized r/AmItheAsshole for the development of machine learning systems that classify or generate ethical decisions.


\section{Dataset}

To answer our research questions, it was necessary to obtain data from two of the popular subreddits.
First, we collected a sample of posts from two popular advice forums on Reddit: r/AmItheAsshole and r/relationships. We used the Reddit API to scrape the top 1,000 posts in the month prior to the collection date from each subreddit~\cite{briggs2020}. 
In all, we collected 989 posts from r/AmItheAsshole (AITA) and 977 from r/relationships. Features collected include unique post ID, created timestamp, post title and text, gilded count, ``flair'' label, number of comments, overall score, and upvote ratio. In Table \ref{table: summary}, we provide a brief summary of the features for each of the two subreddits.

\begin{table}[!htpb]
\centering
\begin{tabular}{p{3.5cm}p{1.25cm}p{1.75cm}}
\toprule
Feature  & AITA & Relationships \\
\toprule
Total ``Reddit gold''  & 52 & 5 \\
Comment count  & 539.53 & 34.38 \\
Score & 4430.38 & 86.58 \\Upvote ratio & 0.96 & 0.83 \\
Word count & 461.51 & 516.42 \\
AFINN sentiment & -4.12 & 5.09 \\
AFINN adjusted & -0.97 & 1.05 \\
Vader compound & 0.30 & 0.30 \\
Vader negative sentiment & 0.09 & 0.09 \\
Vader positive sentiment & 0.12 & 0.12 \\
Cosine similarity & 0.68 & 0.60 \\
Masculine words & 12.99 & 18.18 \\
Feminine words & 18.47 & 12.89 \\
Median age & 25 & 25 \\
\bottomrule
\end{tabular}
\caption{Summary of Data Features by Subreddit. (For definitions, see Section 4.4. Numbers represent averages except where otherwise stated.)}
\label{table: summary}
\end{table}

We processed the text and used a variety of algorithms to generate scores and counts to add to the dataset. Columns added to the dataset included word count, sentiment (VADER positive, negative, and compound as well as AFINN score), cosine similarity, a count of instances of ``masculine'' and ``feminine'' words in each post, and the age and gender of the original poster (when available). Each of these will be explored later in the paper, as the text analysis process itself creates interesting pieces of analysis.

\section{Methodology}

In this section, we explain how we generated and used quantitative data and encoded labels to perform regression analysis and clustering. Our analysis, as explained in the following subsections, helped us to identify patterns in the data.

\subsection{Text Analysis}
Analyzing Reddit posts without performing text analysis would be very limiting, since most of what makes Reddit interesting is in the rich, unstructured writing that often reflects the social norms and rules of each subreddit's distinct community. Because we were looking at advice forums in particular, there was a tremendous amount of text to process and little variety in terms of post type (i.e., no link-only or photo posts, as might be found on other subreddits). The text analysis processes described here address RQ1 -- exploring the top 1000 posts from r/AmItheAsshole and r/relationships by computing quantitative values from the two text corpora.
Pre-processing varied by task, but typically involved converting text to lower case, removing extraneous characters, and in some cases stemming, lemmatization, and stopword removal. We added a column to the dataset that concatenated the post title and the body so all relevant text could be easily accessed from a single document.

\subsubsection{Sentiment}

There is a wide range of sentiment dictionaries available for use with Python. For this research, we made use of two, AFINN and Vader, according to a process laid out by Caren (2019)~\cite{caren2019}.

\paragraph{AFINN}
AFINN analyzes preprocessed text by comparing it to a list of around 2,400 words, each with positive or negative integer values, and applying a total score to each document in a dataset. Because this number is affected by the length of the document itself and the word count of each post varies considerably, we further calculated an ``adjusted AFINN'' by dividing the original AFINN score by the document word count.

The scores themselves are not necessarily that interesting, but the summary statistics for different flair values are somewhat enlightening. For example, r/AmItheAsshole posts judged ``NTA'' or ``not the a-hole''  have an overall lower AFINN score than those judged "YTA" or ``you're the asshole.'' 
Among r/relationships posts, which have a wider range of flair, ``infidelity'' has a much lower average and minimum score than the rest.

\paragraph{Vader (Valence Aware Dictionary and sEntiment Reasoner)}
Vader differs from AFINN in that it analyzes the whole document, not just individual words. Therefore, context and emphasis are considered by Vader. The polarity scores reflect the proportion of a document that is positive, negative, and neutral, and there is also a combined score~\cite{caren2019}. Again, these scores may not be much interesting on their own, but illustrate the variability of the data.

\subsection{Demographics}
The nature of both the forums suggests a lot of gender politics are at work. Digging into that is beyond the scope of this paper, but we are interested in trying to extract and quantify gender and other demographic features from these unstructured text documents that nonetheless follow some conventions around expressing identity.

\subsubsection{Extracting Demographics}

Because posters often follow a similar format to post their age and gender (a 34-year-old woman might write her age and gender as ``(34F)'' or ``'[34f],''
), it is possible to extract demographic information to gain a better understanding of the people who post on these subreddits. We used the Regular Expressions (re) Python library to extract information that followed the pattern ``\emph{number, number, letter}'' or ``\emph{letter, number, number}'' plus the three characters before this pattern (we did not account for posters under the age of 10 because Reddit does not allow anybody under 13 on the platform). Next, we searched for instances of ``\emph{i},'' ``\emph{me},'' or ``\emph{my}'' which would imply that the poster was discussing their own age and gender. Overall, we were able to extract demographic information for 653 posts using this method. 
A manual review of the dataset was conducted to ensure accuracy and fill in demographics data the script missed. The review more than doubled the coverage of demographics data, though many posts did not include this information (92 posts in r/relationships and 356 in r/AmItheAsshole). 

\begin{figure}[h]
\centering
\includegraphics[width=0.99\linewidth]{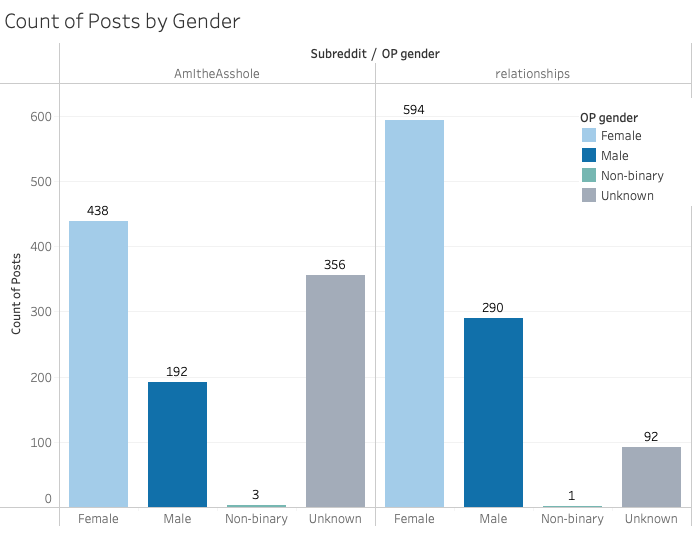}
\caption{Post Count By Identified Gender.}
\label{figure:1}
\end{figure}
\begin{figure}[h]
\centering
\includegraphics[width=0.9\linewidth]{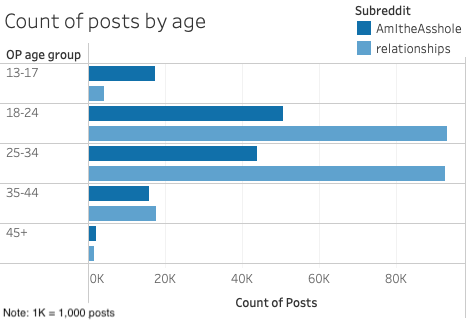}
\caption{Post Count By Identified Ages.}
\label{figure:2}
\end{figure}

Figures \ref{figure:1} and \ref{figure:2} show the post count per subreddit by identified gender and age. In Figure \ref{figure:2}, the ages have been grouped by range and all unknowns have been excluded from the analysis. 
While the number of unknown genders reduce our sample size for the study, we prefer to acknowledge the complexity of gender -- in terms of identity, disclosure, and language -- instead of assuming a binary gender system.
Gender and age were added to the dataset only if they were explicitly stated in the post, and did not include age that were described as a range, such as ``20s'' or ``30s.'' We avoided coding genders for posts that one might \textit{assume} were written by a binary cisgender heterosexual man or woman because although we observed a very low number of non-binary  posters in our dataset, we cannot be certain that is representative of the user community. We did not, for example, assume that just because a person referred to their \textit{husband} they were automatically a \textit{wife} (even without getting into creative uses of such language within the gender expansive community). Unless the poster explicitly identified themselves or used clearly gendered language to refer to themselves, their gender remains unknown.

\subsubsection{‘Masculine’ and ‘Feminine’ Words}

Overall, r/relationships had 90\% coverage for poster demographics, while r/AmItheAsshole only had 64\% coverage. Moreover, the manual review process made some of the differences in identity disclosure practices between the subreddits obvious. R/relationships posters nearly always shared their age and gender, as well as the age(s) and gender(s) of people they write about, in a semi-structured format near the beginning of a post. R/AmItheAsshole posters, however, did not consistently share this information, and did not always use the standard format seen in other forums. In fact, our review identified \textit{only} gender \textit{or} age for 138 posts in that subreddit, compared with only 45 partial demographics in r/relationships. Curiously, while r/AmItheAsshole posts were less likely to explicitly disclose the poster's identity, they routinely state the ages and genders of people the post discussed. The poster's ages ranged from 13 to 60. The gender breakdown can be seen in Figure \ref{figure:1}.

Caren (2019)~\cite{caren2019} described a process for making use of custom and specialized word lists, with ``male'' and ``female'' words as an example. While the original source of those lists, a browser add-on that ``swaps'' gendered words on websites, included hundreds of words to accommodate a wide range of content, our use case is more narrow, so simplified, brief lists of ``masculine'' and ``feminine'' words were created to suit our purposes, primarily covering gendered words for related people advice-seekers are likely to post about, such as \emph{mother} and \emph{father}, \emph{girlfriend} and \emph{boyfriend}, \emph{wife} and \emph{husband}. We considered whether ``gender neutral'' word lists could also be created to include gender-diverse posters, but decided against it, as it seemed likely to result in false positives.


\subsection{Similarity}

There are many vector space modeling tools that calculate cosine similarity scores for a set of text documents. We wanted one that would help us explore both similarity and uniqueness 
and let us add a single score to our dataframe for further analysis. Doc2vec compares documents (which could also be longer posts or descriptions) and assesses the similarity between them on a n-dimensional space. The use of Doc2Vec has been common for similar tasks such as identifying unique movie plot descriptions~\cite{lau2016empirical, nag2019}.

First, we transformed the corpus into document vectors, then trained an auto-encoder neural network to predict the vectors. The auto-encoder accuracy score takes into account both predicted and actual vectors, with any result above 0.95 being excellent. The scores for each subreddit surpassed this threshold.
Further, we generated the top five most unique posts for each subreddit. To get the singular similarity score and apply it to each document, we calculated the cosine similarity between input vector and the output vector of $i^{th}$ data point.

To better understand the content of the posts in r/relationships and r/AmITheAsshole, we extracted a list of post titles, parsed the text into individual tokens, and examined the nouns and verbs they contained. Then we classified and applied parts of speech tagging using the Python NLTK library. To reduce token redundancy, we normalized verb and noun tokens, then filtered 
by base word to group variants, and sorted by frequency.

\subsection{Feature Generation} 
Features 1-4 in the following list were generated directly from the Reddit API. We define these Reddit features as interaction metrics that may be a predictable outcome. The rest of the features -- numeric, and required for statistical analysis -- were generated using the text analysis processes described in the previous section, and can be described as follows: 

\noindent
1)~\textbf{gilded}: numeric value between 0 and 4 denoting the number of paid ``gold'' stars that were given to a particular post.

\noindent
2)~\textbf{num\_comments}: numeric value between 0 and 6,821 indicating the number of comments on a post.

\noindent
3)~\textbf{score}: numeric value between 1 and 35,188, representing the total number of upvotes minus the total number of downvotes. 

\noindent
4)~\textbf{upvote\_ratio}: numeric value between 0.55 and 1 representing the ratio of upvotes vs. downvotes.

\noindent
5)~\textbf{afinn\_score}: calculated numeric value between -98 and 133 representing a measure of sentiment analysis.

\noindent
6)~\textbf{word\_count}: calculated numeric value between 74 and 6,494.

\noindent
7)~\textbf{afinn\_adjusted}: calculated numeric value between -22.047244 and 27.108434 that accounts for word count.

\noindent
8)~\textbf{vader\_compound}: calculated numeric value between -0.9994 and 0.9999 representing a normalized, weighted composite sentiment score.

\noindent
9)~\textbf{vader\_neg}: calculated numeric value between 0 and 0.293 representing the proportion of a text that has a ``negative'' sentiment score.

\noindent
10)~\textbf{vader\_pos}: calculated numeric value between 0.011 and 0.29 representing the proportion of the text that has a ``positive'' sentiment score.

\noindent
11)~\textbf{masc\_words}: calculated numeric value between 0 and 210 based on a list of ``masculine'' words created for this task.

\noindent
12)~\textbf{fem\_words}: calculated numeric value between 0 and 337 based on a list of ``feminine'' words created for this task.

\noindent
13)~\textbf{cosine\_similarity}: calculated numeric value between -0.167519 and 0.99984 based on Doc2Vec.

\noindent
14)~\textbf{OP\_demographics}: the age and gender of the poster, formatted as ``age,gender''.

\noindent
15)~\textbf{OP\_age}: the age of the poster, expressed as a number.

\noindent
16)~\textbf{OP\_gender}: the gender of the poster, expressed as either m (male), f (female), n (nonbinary), or unknown.

\section{Results and Discussion}

In the following subsections, we discuss our results. 

\subsection{Sentiment Polarity}

To answer RQ1, we 
extracted several natural language features from Reddit posts. We captured each poster's age and gender whenever possible. Overall word count, as well as a count of how many matches against our ``gendered word lists'' existed for masculine and feminine words, are simple calculations. To quantify sentiment, we used two algorithms, AFINN and Vader, to generate a total of five numeric features, AFINN score, AFINN adjusted for post length, Vader compound, Vader positive, and Vader negative. To explore language and similarity, we generated a cosine similarity score for each post, and performed additional text processing to determine how often words were used in each subreddit, including part-of-speech analysis.

Our comparison of the two sentiment detection algorithms -- AFINN and Vader -- show that for this dataset, AFINN (adjusted for document length) and Vader compound disagree on the positivity or negativity about 35\% of the time, with Vader being more than twice as likely to consider a post to have a positive sentiment when AFINN calculates it as having a negative sentiment. 

\subsection{Subreddit Comparison}

To answer RQ2, we compared the subreddits in terms of length, demographics, sentiment, commonly used language, and gender. These features, paired with qualitative explorations of each subreddit and and its community rules, enabled us to make observations about the nature of personal disclosures and relationships of interest for users of these forums.

Broadly speaking, r/relationships posts are, on average, 55 words longer than the average r/AmItheAsshole post, and more than 90\% of those include semi-structured demographic disclosures compared to just 64\% of r/AmItheAsshole posts. While manually reviewing the data for age/gender disclosures missed by the automated extraction, we observed a marked difference in each community's norms around such information which other researchers such as Botzer, et. al. (2021) [4] had also observed. Posters on both subreddits routinely shared identity markers about the subjects of their posts (e.g., wife, father-in-law, children), but many r/AmItheAsshole posts failed to explicitly identify themselves in the same way. In some cases, their identities were stated as part of the post narrative or in an edited comment added to the post body in response to community feedback (``I'm a woman, not a man''). There were also more partial disclosures, that is, only age or gender (see Table \ref{table:5}).

\begin{table}[h]
\begin{tabular}{p{2.65cm}p{1.7cm}p{2.15cm}}
\toprule
 & r/relationships & r/AmItheAsshole \\
 \toprule
Both age \& gender & 866 & 516 \\
Only age & 19 & 21 \\
Only gender & 26 & 117 \\
\bottomrule
\end{tabular}
\caption{Demographics Disclosure for Each Subreddit.}
\label{table:5}
\end{table}

When comparing the rules dictating participation and moderation on each subreddit, the reason becomes clearer. One requirement of r/relationships is that each post much include a ``TL;DR'' (too long; didn't read) summarizing their ask, which could account for the difference in average length. The primary rule on r/relationships is that ages, genders, and relationship length are required components of any post, which is not the case on the other forum, where (humorously enough) the first rule is ``Be civil.'' Curiously, r/AmItheAsshole's strictly enforced rules describe it as ``NOT an advice sub,'' though in the context of this analysis we'd argue that asking both what one \textit{should do} and what one \textit{may have done wrong} have behavioral implications for human interaction.

The subreddits' distinctive rules and structure also affect how we might compare the two on the question of flair. On r/AmItheAsshole, flair is used by moderators to label judgment: ``ass'' for predominantly ``YTA'' (you're the a-hole) responses and ``not'' for ``NTA'' (not the a-hole). Other judgments exist on the site that are not found in this dataset, reflecting situations like ``no one's the a-hole'' and ``everyone's an a-hole.'' Posts without a final verdict have no flair at all. R/relationships uses flair to classify posts by topic, predominantly ``Relationships'' and ``Updates,'' but also ``Non-Romantic,'' ``Infidelity,'' ``Breakups,'' and so on. Posters assign their own flair after submitting a post. Because of these major differences in the nature of each subreddit's flair, we were limited in how effectively we can compare them on a feature level. For example, when we looked at whether there are gender differences among flair labels, only a few outliers emerged. 45\% of unknown gender posts in r/relationships have a ``new'' label, and may be edited post-submission to comply with the rules. On r/AmItheAsshole, where ``NTA'' outnumbers ``YTA'' 8-to-1 and there are more than twice as many female posters than male, men are disproportionately likely to receive an ``you're the a-hole'' judgment (17.7\% of men's posts vs. just 6.8\% of women's).

A minor difference between the subreddits was observed regarding posts that explicitly called out their use of alternative profiles for the purpose of anonymity. 48 r/AmItheAsshole posts used ``throwaway'' or ``anon'' compared with 27 r/relationships posts. This represents too small a sample in our dataset for deeper analysis, but could be explored in future with targeted sampling methods.

\subsubsection{Sentiment}

We calculated sentiment using two different models, AFINN and Vader. In aggregate, Vader showed few subreddit-level differences, but AFINN did. Looking at AFINN scores (adjusted to account for the effect of document length), r/relationships have a wider range of both positive and negative scores compared with r/AmItheAsshole, and its average score is slightly positive instead of slightly negative (see Table \ref{table:6}).
There were some outliers when we further analyzed AFINN by flair and gender, but they were unlikely to be significant, as they represent relatively low-frequency groups (less than 1\% of the dataset): Female r/AmItheAsshole posts with no flair, male update posts in r/relationships, and infidelity posts by posters whose gender is unknown. Overall, sentiment does not offer much in the way of useful differentiation between the subreddits.

\begin{table}[h]
\begin{tabular}{llrr}
\toprule
\multicolumn{1}{l}{} & & r/AmItheAsshole & \multicolumn{1}{r}{r/relationships} \\ 
\toprule
\multirow{3}{*}{AFINN} & Avg & -0.97 & 1.05 \\
 & Min & -13.99 & -22.05 \\
 & Max & 12.36 & 27.11 \\ 
 \midrule
\multirow{3}{*}{Vader} & Avg & 0.91 & 0.92 \\
 & Min & -1.00 & -1.00 \\
 & Max & 1.00 & 1.00 \\ 
 \bottomrule
\end{tabular}
\caption{Comparison of Adjusted AFINN and Vader Compound Sentiment Scores for Each Subreddit.}
\label{table:6}
\end{table}

\subsubsection{Language}

We compared these subreddits on a word frequency level to get a sense of whether common themes emerged. Looking at just verbs
, we found that r/AmItheAsshole post titles disproportionately use verbs like  \textit{tell}, \textit{want}, \textit{get}, \textit{refuse}, \textit{give}, and \textit{let}, while r/relationships posters are more likely to \textit{be}, \textit{do}, \textit{feel}, and \textit{know}. It makes sense that users of the former's arguably more ``aggressive'' forum might be more interested in what they (or others) desire or say, while the latter demonstrates more interiority. Switching to nouns
, we see prominent use of (often gendered) words for relations, but interestingly, r/AmItheAsshole posts tend toward family terms like \textit{mother}, \textit{sister}, \textit{daughter}, \textit{husband}, \textit{father}, \textit{brother}, \textit{family}, \textit{wife}, \textit{kid}, \textit{son}, and \textit{parent}, while the advice-seeking r/relationships posters use more  \textit{I}, \textit{boyfriend}, \textit{girlfriend}, \textit{relationship}, and \textit{partner}. 

\begin{table*}[h]
\centering
\begin{tabular}{ll}
\toprule
Post Title & Cosine Sim Val \\
\toprule
\multicolumn{2}{l}{r/AmItheAsshole Subreddit} \\
\cmidrule(l){1-1}
AITA for telling my mum that she is creating a future drug addict & -0.09337831    \\
\hline
\begin{tabular}[c]{@{}l@{}}AITA for telling my SIL who is a professional chef to stop being \\ a loudmouth after she decided to be a backseat cook in my kitchen?\end{tabular}    & -0.11074713    \\
\hline
AITA for not learning my husband’s native language? & -0.11149485    \\
\hline
AITA for correcting people who assume my stepmom is my mom? & -0.11463856    \\
\hline
AITA for not giving a bully’s mom a promotion? & -0.12220774    \\
\bottomrule
\multicolumn{2}{l}{r/relationships Subreddit} \\
\cmidrule(l){1-1}
\begin{tabular}[c]{@{}l@{}}I 28M when on a date with 30F, seeking communication advice \\ between dates\end{tabular} & -0.18039174    \\
\hline
My(22F) boyfriend (28M) only tells me he loves me during sex & -0.14639542    \\
\hline
\begin{tabular}[c]{@{}l@{}}My(32m)gf(f30) ex husband(m?) is back with the girl he left her for \\ and I think it bothers her\end{tabular} & -0.13790806    \\
\hline
Guys who's gf has ''he's just a friend'' friend & -0.09033157    \\
\hline
\begin{tabular}[c]{@{}l@{}}I (20F) had a crush on a now unavailable friend (21M), who (along \\ with his new GF) is now asking me what's wrong. What do I tell them?\end{tabular} & -0.08882173\\
\hline
\end{tabular}
\caption{Titles of Top Five ``Most Unique'' Posts in Each Subreddit According to Cosine Similarity Score.}
\label{table:uniqueposts}
\end{table*}

In addition to most of these nouns being gendered, they also suggest some clear divergence in the nature of these forums in terms of the information posters seek as well as disclose about themselves.
Cosine similarity enabled us to compare the subreddits at the word level, and additionally, to explore uniqueness at the post level. 
Table \ref{table:uniqueposts} shows the post titles for the top five most unique entries in both subreddits.

\subsubsection{Genderedness}


When we dig into expressed gender identity and gendered word use, we find that in general these subreddits are broadly, but not exclusively, cisheteronormative. Just four posts could be attributed to gender diverse individuals; while we did not call out posts that described a binary transgender experience, we observed that this is not a common theme. In r/AmItheAsshole especially, where a large percentage of posters do not explicitly state their own gender identity, there may be a presumption of gender from context if we assume heteronormativity (which we did not). Differentiating between those posts where the user deliberately concealed that aspect of their identity from those where the user assumed they would be read as their assigned gender, based on a provided description of their relationship(s) with others, could be a project unto itself. Users may make these assumptions, but careful researchers should not.
For those posts where we \textit{can} safely account for gender, we explored the relationship between gender and gendered language for these subreddits:


\noindent
\textbf{1)}~While women's r/relationships posts used gendered language more than twice as often as men's, women's used almost three times as many masculine words as feminine. On the other hand, men's posts had a similar ratio of feminine words to masculine. 

\noindent
\textbf{2)}~Similarly, r/AmItheAsshole posts by men used gendered language half as much as women's and men used twice as many feminine words than masculine. Women's posts used only slightly more masculine words than feminine.

\noindent
\textbf{3)}~Women's posts are, on average, 12 words longer than men's on r/relationships, and 35 words longer on r/AmItheAsshole. Posts with an identified gender are around 30 words longer than those without on both subreddits.

\noindent
\textbf{4)}~On r/relationships, women's posts garner more comments on average than men's (36 vs. 29 comments), but posts where the gender is unknown, the average is higher than both (42 comments). On r/AmItheAsshole, men's posts have more comments on average (602 vs. 547 for women and 499 for unknown).

\noindent
\textbf{5)}~Looking at average sentiment, r/relationships posts by men are marginally more ``positive'' than those by women according to both AFINN (adjusted; 1.4 vs. 0.95) and Vader (compound; 0.36 vs. 0.27). The opposite is somewhat true for r/AmItheAsshole: women's AFINN adjusted scores are less negative than men's (-0.74 vs. -1.23) and Vader compound scores more positive (0.33 vs. 0.24). That these models disagree on positivity/negativity suggests there may be some language used more often in r/AmItheAsshole posts that's computed as negative in only one model (like the word ``asshole,'' for example).


The nature of r/AmItheAsshole's flair assignments gives us the opportunity to explore the relationship between gender and crowdsourced behavioral judgments (i.e., who's the a-hole). While ``YTA'' posts are undersampled in our dataset and women outnumber men, men are more likely to receive the ``a-hole'' judgment: 17.7\% of men's posts vs. 6.8\% of women's. But in terms of total ``YTA'' posts, a plurality are earned by those with an unknown gender. To explore this difference further, we would need to collect more data that oversamples for less-common flair values, including ``YTA.''

\subsection{Engagement Metrics}

To answer RQ3, we investigated whether there were any correlations between natural language features and post engagement metrics in our dataset.
First, we compared engagement across subreddits. R/relationships posts received far fewer comments -- 34, on average, compared with r/AmItheAsshole's 540. In fact, of our sample, r/AmItheAsshole post engagement metrics towered over r/relationships overall. This was reflected in the score, a straightforward measure of total upvotes minus downvotes. Despite scraping the ``top 1,000'' posts in each subreddit for the same time period and at the same time, r/relationships includes at least one post with no comments at all and a score of just one. The \textit{minimum} score of an r/AmItheAsshole post in this dataset is more than seven times higher than the \textit{average} score for r/relationships (673 vs. 87). One might attribute an inflated upvote ratio on r/AmItheAsshole to the community rule against ``downvoting a-holes,'' but the score demonstrates overall higher engagement not fully explained by that community standard. Other researchers 
have offered explanations for this upvote phenomena, such as that dissent is affectively discouraged in subreddit communities because ``defection from normative community values (as indicated by downvotes) harbors emotional consequences and attention rewards'' (Davis et al., 2021, p. 13~\cite{davis2021}). Higher comment counts could also be related to subreddit rules: debate is banned on r/AmItheAsshole while judgment is swift, lowering the bar for contribution compared with r/relationships, which may require a more thoughtful response than ``NTA.'' Although r/relationships has more subscribers, it's r/AmItheAsshole posts that tend to go ``viral,'' perhaps attracting engagement from Reddit users who do not regularly participate in the community. However, this dataset includes neither post username information nor comments, so the question of virality and the nature of community participation is best left for future study.

However, if we look at how much engagement metrics correlate with the computed and extracted language features in our dataset, the relationship is very weak at best
, both overall and for each subreddit.
Although the correlation coefficients are low overall, we were particularly interested in the impact of uniqueness, as represented by low cosine similarity. If we graph cosine similarity scores, it is clear that a vast majority of posts are very similar, 0.50 or greater. There are 138 posts with negative similarity scores and 555 if we include all posts with a score of 0.25 or less. 
There are slight differences between the subreddits in terms of correlation between similarity and engagement.

\section{Conclusions and Future Work}

In this work, we developed a methodology to extract demographic information from a social media site where user demographics are neither always  shared, nor presented in a structured format. We investigated what natural language features could be extracted from unstructured Reddit post text, compared two advice-oriented subreddits in terms of those features, and looked for correlations between language features and engagement metrics. The features we extracted included sentiment, similarity, word frequency, and in-text demographics. 
We examined how the post authors’ self-disclosures  reflect the unique communities in which their posts are shared and how the authors’ language use choices might be related to broader social patterns. 
We found some differences between r/AmItheAsshole and r/relationships, especially in terms of word frequency, demographics disclosure, and gendered language, but nothing significant in terms of sentiment or overall similarity. A more in-depth analysis of parts-of-speech tagging seems like the best path forward in terms of text analysis. 
We obtained the most common verbs and nouns found in the top 1000 posts of each of the r/relationships and r/AmItheAsshole subreddit communities, which could be a practical tool to help the scholarly community identify and compare high frequency words in other advice forums in future studies. Our observations on the differences in aggression and interiority of different subreddits could inspire future studies to explore how posters' perceived locus of control and relational depth influences their query formulation. 
While these topics fall outside the scope of this study, future researchers could examine additional online advice communities, improve demographics extraction methods, and perform more detailed parts of speech tagging and token analyses of posts’ text to look for patterns.

This study is limited in that it relies on a one-time scraping of ``top'' Reddit posts that may be biased in favor of higher engagement. It does not attempt to account for time or over-sample for  certain areas of interest that are unbalanced, like flair and gender diversity. We did not collect comments or user metadata, so it cannot address a broader look at interactions and networks on Reddit. Future work could take these into account,  apply more robust processes to extract demographic information from posts, and  identify coding methods for key words and phrases for deeper analysis. 
We would also love to explore the relationship type between the original poster (OP) and subject without relying on assumptions about the gender binary.
The scope could be broadened to include other relationship- and behavior-centric forums to explore how different communities create and share information about ethical decision-making and human behavior. Another aspect to explore is the ``memeification'' of the posts in these forums, that is, how posts in some subreddits are captured and shared on other social media networks, which may further reflect cultural values and acceptable behaviors.



\bibliographystyle{ieeetr}
\bibliography{hicss-2022}

\end{document}